\documentclass[screen,nonacm]{acmart}
\newcommand{\xhdr}[1]{\vspace{1.7mm}\noindent{{\bf #1.}}}

\usepackage{wrapfig} 
\usepackage{enumitem}
\usepackage{adjustbox}

\newcommand{\eg}{\textit{e.g.,}}
\newcommand{\ie}{\textit{i.e.,}}

\AtBeginDocument{%
  }

\setcopyright{none}

\begin{document}

\title[Disclosure or Marketing?]{Disclosure or Marketing? Analyzing the Efficacy of Vendor Self-reports for Vetting Public-sector AI}

\author{Blaine Kuehnert}
\orcid{0009-0001-5702-5693}
\authornote{Both authors contributed equally to this research.}
\affiliation{%
  \institution{Carnegie Mellon University}
  \city{Pittsburgh}
  \state{Pennsylvania}
  \country{USA}
}
\email{blainekuehnert@cmu.edu}

\author{Nari Johnson}
\orcid{0009-0008-3180-3582}
\authornotemark[1]
\affiliation{%
  \institution{Carnegie Mellon University}
  \city{Pittsburgh}
  \state{Pennsylvania}
  \country{USA}
}
\email{narij@andrew.cmu.edu}

\author{Ravit Dotan}
\orcid{0000-0002-9646-8315}
\affiliation{%
  \institution{TechBetter}
  \city{Pittsburgh}
  \state{Pennsylvania}
  \country{USA}
}
\email{ravit@techbetter.ai}

\author{Hoda Heidari}
\orcid{0000-0003-3710-4076}
\affiliation{%
  \institution{Carnegie Mellon University}
  \city{Pittsburgh}
  \state{Pennsylvania}
  \country{USA}
}
\email{hheidari@andrew.cmu.edu}

\renewcommand{\shortauthors}{Kuehnert et al.}

\begin{abstract}
  Documentation-based disclosure has become a central governance strategy for responsible AI, particularly in public-sector procurement. Tools such as model cards, datasheets, and AI FactSheets are increasingly expected to support accountability, risk assessment, and informed decision-making across organizational boundaries. Yet there is limited empirical evidence about how these artifacts are produced, interpreted, and used in practice. In this paper, we present a qualitative study of the GovAI Coalition FactSheet, a widely adopted transparency document designed to support AI procurement and governance in government contexts. Drawing on semi-structured interviews with vendors and public-sector practitioners, alongside a systematic analysis of completed FactSheets, we examine how FactSheets are used, what information they surface, and where they fall short. We find that FactSheets are asked to serve multiple and conflicting purposes simultaneously: showcasing vendor offerings, supporting evaluation and due diligence, and facilitating early-stage dialogue between vendors and agencies. These competing expectations, combined with the structural constraints of voluntary and public self-disclosure, limit the ability of FactSheets to function as standalone evaluation or risk-assessment tools. At the same time, our findings suggest that when understood as relational artifacts used to establish trust, shared understanding, and ongoing dialogue, FactSheets can help create conditions that support more meaningful disclosure and governance over time.
\end{abstract}




\maketitle

\section{Introduction}
Governments across the world are increasingly adopting Artificial Intelligence (``AI'') systems, with significant implications for the rights and safety of residents \citep{reisman2018algorithmic,kim2024public,chouldechova2018case,eubanks2018automating,whitney2021hci,kawakami2024studyingpublicsectorai,levy2021algorithms}. Within the United States, FAccT scholars have examined how algorithmic design decisions embed value judgments and shape policy outcomes in domains such as social services \citep{eubanks2018automating,kawakami2022improving,gerchick2023devil} and the criminal legal system \citep{meyer2022flipping,pruss2023ghosting}. Resource constraints in the public sector \citep{johnson2025legacy}, combined with the rapid availability of generative AI tools \citep{chen2025dispelling}, have further intensified demand for AI systems and heightened the urgency of responsible oversight.

Meaningful oversight, however, is rarely straightforward. Most public-sector AI systems are not developed internally, but procured from private vendors \citep{mulligan2019procurement,rubenstein2020federal,hickok2024public}. This reliance on third-party vendors creates structural barriers to governments’ ability to understand and govern the systems they deploy, including claims of corporate secrecy \citep{pasquale2015black,brauneis2018algorithmic}, limited access to technical expertise \citep{engstrom2021artificially}, lack of participation in system design \citep{veale2018fairness}, and the intrinsic complexity of modern AI systems \citep{lipton2017mythos,liao2024transparency}. These challenges are especially acute for under-resourced local governments \citep{johnson2025legacy}, producing persistent information asymmetries that complicate risk assessment, accountability, and decision-making.
In such contexts, prior literature emphasizes that meaningful transparency emerges through relational interactions between multiple actors like vendors, procuring agencies, and external evaluators, who collectively interpret and act on system information~\citep{cobbe23supplychains, nissenbaum2009privacy}. 
Procurement provides a particularly important lens for studying these dynamics because it is the moment when governments must make concrete decisions about adopting AI systems while relying largely on information provided by vendors ~\citep{brauneis2018algorithmic,mulligan2019procurement,kawakami2024studyingpublicsectorai,johnson2025legacy}. 
Documentation artifacts such as FactSheets are therefore intended to reduce information asymmetries at precisely the stage when they are most consequential.

In response, scholars and advocates have called for governments to play a more active role in interrogating AI systems during procurement \citep{mulligan2019procurement,richardson2019confronting,rubenstein2021acquiring,hickok2024public}. A wide range of documentation-based artifacts have emerged to support this goal, often in the form of structured question sets designed to elicit disclosures about the practical capabilities, technical aspects, and potential harms of AI systems from vendors \citep{wefbox,conti-cookguiding,dotan2023how}. 
These artifacts are notable because they rely on voluntary self-disclosure by vendors, often in publicly accessible formats, making them a lightweight and scalable governance mechanism ~\citep{heger2022understanding,tang2025navigating}.
One prominent example is the \emph{AI FactSheet} \citep{govai_factsheet}, developed by the GovAI Coalition: a collaborative of U.S. state and local governments seeking practical tools for AI governance \citep{govai_openletter}. 
The GovAI Factsheet builds on earlier documentation initiatives in responsible AI, such as model cards ~\citep{mitchell2019model} or IBM's original AI factsheet for AI developers ~\citep{arnold2019factsheet}, while adapting the concept to the unique context of public-sector procurement. 
The GovAI FactSheet consists of 23 open-ended questions that invite vendors to disclose information about system design, data, evaluation, and intended use. Over the past two years, it has become widely adopted, with hundreds of vendors and more than 300 public-sector organizations using the tool in practice \citep{sharma2024power}. As such, the FactSheet offers a valuable case study of documentation-based approaches that now sit at the center of many AI governance regimes, including model cards \citep{mitchell2019model}, transparency reports \citep{bommasani2024foundation}, and algorithmic impact assessments \citep{moss2021assembling}.

Despite their growing adoption, there has been little empirical examination of whether these self-disclosure artifacts achieve the goals advocates and scholars have articulated for them—namely, enabling public servants to meaningfully understand, assess, and govern AI systems. More broadly, we lack clarity about how such documentation is used in practice: what governments hope to achieve by requesting it, how vendors approach completing it, and what kinds of information are ultimately disclosed. Addressing this gap is essential for evaluating whether documentation fulfills its intended governance role or whether structural constraints limit what it can realistically deliver. We therefore ask: 
\begin{itemize}[leftmargin=1em,labelsep=0.5em]
\item \textbf{RQ1:} What technical, evaluative, and governance-related \textbf{information} do vendors elect to voluntarily disclose, and what information is consistently limited or omitted?
\item \textbf{RQ2:} What \textbf{goals} do public sector agencies and AI vendors have when requesting or completing AI disclosure documentation? How do these stakeholders make use of the completed FactSheets?
\item \textbf{RQ3:} What \textbf{broader conditions} and challenges shape information disclosures about public sector AI systems?
\end{itemize}
To answer these questions, we conduct a mixed-methods empirical study of the GovAI FactSheet. We draw on over 20 hours of qualitative interviews with AI vendors and public-sector practitioners, analysis of 39 completed FactSheets for AI systems sold to U.S. local governments, and months of embedded participant observation within the GovAI Coalition. Together, these data allow us to examine both how stakeholders conceptualize the purpose of documentation and what kinds of disclosures it produces in practice.

Our analysis reveals that FactSheets struggle to fulfill the governance roles often expected of them, largely because different stakeholders approach these artifacts with competing goals. Governments frequently seek documentation to support procurement due diligence, risk assessment, or compliance, while vendors often view FactSheets as marketing tools or low-risk obligations. These differing orientations create a mismatch between what governments expect documentation to provide and what vendors feel able or incentivized to disclose, resulting in disclosures that often lack the system design, evaluation, or governance detail needed to support public-sector decision making.

Rather than treating this gap as a failure of documentation alone, our findings suggest a fundamentally different way to understand the role of vendor self-disclosure. Vendors’ willingness to share information is shaped by legal risk, competitive pressure, uncertainty about interpretation, and limited incentives for detailed disclosure.
In this context, FactSheets frequently fall short as standalone assessment tools, but nonetheless play an important role in structuring early-stage dialogue and establishing shared understanding. We argue that meaningful assessment is more likely to emerge through layered governance processes in which documentation serves as an entry point, complemented by interactive, empirical, or third-party evaluation mechanisms. 
 
In doing so, this paper contributes a critical perspective on documentation-based AI governance, highlighting the relational dynamics through which transparency is negotiated and interpreted between vendors, governments, and other stakeholders, and the institutional constraints that shape what transparency can achieve in practice. 
We conclude by outlining implications for researchers and policymakers designing disclosure mechanisms that better align with procurement realities, including clearer purpose alignment, differentiated audiences, and incentive structures that reward meaningful disclosure rather than superficial compliance.
\vspace{-5pt}
\section{Background}
\xhdr{AI Use and Procurement in Government} FAccT researchers have long examined the usage of AI systems within the public sector, recognizing that local governments oversee many of the most critical social services that impact historically marginalized populations, such as child welfare risk assessment ~\cite{gerchick2023devil, stapleton2022imagining, saxena2021a, saxena2020government}, policing~\cite{ziosi2024evidence, alikhademi2022review}, and adjudicating criminal legal cases~\cite{wang2024against, yacoby2022if, meyer2022flipping}. In the United States, where our study is situated, past work has shown how public-sector AI systems encode values, reshape discretion, and redistribute accountability across public sector workers, agency leadership, and AI vendors \cite{crump2016surveillance,mulligan2019procurement,whitney2021hci,gerchick2023devil}.
Our study takes place amidst a rapid expansion in the availability of generative AI systems framed as promoting ``government efficiency'' \cite{chen2025dispelling,anexreis2025deploy}, such as copilots for frontline workers \cite{eu_chatgpt_report,pa_chatgpt_report}, or other emerging use cases, such as chatbots that can automate service provision to residents \cite{engstrom2021artificially}.

However, research has demonstrated that public agencies often lack meaningful visibility into the AI systems they procure \cite{veale2018fairness,eubanks2018automating,brauneis2018algorithmic,mulligan2019procurement,johnson2025legacy}. In their analysis of open records requests made of forty-two U.S. agencies, \citet{brauneis2018algorithmic} found that governments consistently had no access to basic information about the AI systems they were using, such as knowledge of what data they were trained on, or the input features used to make decisions. This opacity is especially concerning when AI systems displace or constrain the discretion of public servants, who remain legally and ethically accountable for the decisions shaped by often-inscrutable AI systems \cite{pasquale2015black,reisman2018algorithmic,alkhatib2019street}. In this sense, \citet{mulligan2019procurement} argue, procured AI systems effectively “make policy'', and greater transparency is a necessary first step to enable meaningful government scrutiny and participation in ensuring that the values encoded by AI design decisions are in line with public values.

In response to these concerns, scholars and civil society groups have increasingly turned to public procurement, defined as ``the formal processes through which governments acquire goods and services from external vendors'' \cite{lloyd2004public}, as a site to pursue AI governance interventions \cite{reisman2018algorithmic,rubenstein2020federal,dotan2023how,hickok2024public,pakzad2025key}. 
Building on longer traditions of procurement as a mechanism to promote social values such as environmental sustainability \cite{lazaroiu2020environment}, many recent AI procurement reforms emphasize transparency through structured documentation and disclosure requirements (\eg{} put forward by groups like the World Economic Forum ~\cite{wefbox} or IEEE ~\cite{ieee_standard}), with the aim of enabling more informed evaluation and oversight of vendor-developed systems. 
Our work contributes to this literature by shifting attention from what governments should ask, to what vendors actually disclose in practice. We present the first empirical analysis of public sector AI vendor disclosures completed using a standardized template: the GovAI factsheet.

\xhdr{The GovAI Coalition and the AI Factsheet}
In response to many of the challenges discussed in the previous section, the GovAI Coalition emerged in 2023 as a collaborative effort among U.S. state and local governments seeking practical tools to govern AI adoption~\cite{govai_openletter}. Motivated by growing interest in Generative AI and mounting concerns about information asymmetry, the coalition issued an open letter announcing its AI Factsheet: a standardized set of 23 open-ended questions designed for public agencies to pose to AI vendors during procurement~\cite{govai_openletter}. The coalition initially had two goals in mind when introducing the FactSheet: (1) enable governments to obtain foundational information about vendor systems, and (2) encourage vendors to voluntarily disclose details that would support responsible deployment, oversight, and risk management for AI in local governments~\cite{govai_openletter}.

The coalition has since grown to over 600 public-sector organizations, with over 50 vendors participating in coalition activities \cite{govai_phasetwo}. The GovAI coalition understood that governments may want to use the FactSheet in multiple ways. Some may want to integrate it directly into procurement requirements, while others may request it during early scoping conversations. The coalition works to support each of these efforts by maintaining a public registry of completed FactSheets, effectively creating a searchable marketplace of AI systems~\cite{govai_openletter}. In practice, completing and submitting the FactSheet has also become an initial requirement to membership for vendors seeking to join the coalition or sell to member agencies.

Despite this widespread adoption, there is limited empirical research evaluating how FactSheets are used, how vendors complete them, or whether they meaningfully support governments’ ability to assess AI systems. Early signals from coalition members suggest that governments often struggle to interpret responses, especially when vendors provide high-level or marketing-oriented answers. GovAI’s current internal review processes focus primarily on completeness rather than evaluating technical quality or accuracy, leaving individual governments to interpret vendor responses in ways that vary widely in ability and usefulness. But this is not surprising. We know from work by \citet{johnson2025legacy} that U.S. government agencies are struggling to interpret vendor's responses on documentation tools like the FactSheet. This gap in understanding motivates our study. The FactSheet is now one of the most widely adopted self-disclosure tools in U.S. state and local government AI procurement, yet little is known about how it is actually used or what kinds of information it produces. Examining this artifact provides a timely and representative case study of the promises and limits of documentation-based self-disclosure within real-world public-sector procurement.

\xhdr{Disclosure Documentation in Responsible AI Governance}
We situate our study of the GovAI FactSheet as one example of a much broader class of documentation artifacts that rely on vendor self-disclosure to support responsible AI governance.
Within the FAccT community, documentation has long been proposed as a mechanism for improving disclosure, accountability, and oversight in AI systems. Early work on documentation in machine learning such as Model Cards for Model Reporting~\cite{mitchell2019model} and Datasheets for Datasets~\cite{Gebru2021} argue that structured documentation could uncover design decisions laden with implicit values, encourage responsible development practices, and provide stakeholders with important information about system behavior, limitations, and appropriate use. 
Similar forms of self-disclosure appear across AI ecosystems, from model cards and usage statements shared on platforms like Hugging Face ~\cite{tang2025navigating}, to transparency reports and compliance-oriented documentation increasingly referenced in policy and regulatory proposals~\cite{EU_AI_Act_2024, nyc_local_law_144}. 
Understanding how vendors approach these disclosures is particularly important given the extent to which disclosure-based interventions are now widely proposed, and in some cases mandated, as a foundational mechanism for AI governance and oversight, both within the public procurement process \cite{wefbox,conti-cookguiding}, and in other regulatory regimes, such as the European Union's AI Act, which requires providers of high-risk AI systems to produce technical documentation describing key system design choices \cite{disclosurebydesign22norval,moss2021assembling,wright2024null}.

Recent empirical scholarship has also begun to examine documentation in practice. 
Interview-based research with practitioners responsible for authoring AI documentation has shown that voluntary disclosures are shaped by both technical limitations and operational considerations, including concerns about competitive disadvantage, legal exposure, and reputational harm~\cite{chang2022understanding, heger2022understanding}.
This body of work highlights how documentation is not simply a neutral reporting mechanism, but a negotiated artifact shaped by stakeholder incentives and institutional context.
Within this emerging literature, our study contributes a focused examination of vendor self-disclosure in the context of public-sector AI procurement, which is quickly becoming an especially consequential setting where documentation is expected to inform purchasing decisions, risk assessment, and public accountability.
\vspace{-5pt}
\section{Methodology}
\label{methodology}
To examine how self-disclosure documentation functions in public-sector AI procurement, we draw on two primary sources of empirical data: semi-structured interviews with government and vendor stakeholders, and systematic analysis of completed AI FactSheets submitted to the GovAI Coalition. Together, these approaches allow us to examine both how disclosure documentation is understood by key actors and how it operates in practice.

\begin{table}[t]
\centering
\footnotesize
\caption{\textbf{Interview participants.} Vendor and public-sector participants included a range of roles and organization types. Titles and agency descriptions were slightly modified to preserve anonymity.}
\label{tab:participants}
\adjustbox{valign=t}{\begin{minipage}{0.45\textwidth}
\centering
\textbf{(a) Vendor Participants}
\begin{tabular}{llll}
\toprule
ID & Application & Role & Org Size \\
\midrule
P1 & AI Governance Platform & All & 1 \\
P2 & Residential Services AI & Consultant & 11--50 \\
P3 & Residential Services AI & Marketing Lead & 11--50 \\
P4 & Proposal Evaluation & CEO & 11--50 \\
P5 & Proposal Evaluation & CTO & 11--50 \\
P6 & Knowledge Assistant & CEO & 11--50 \\
P7 & Survey Data Analysis & VP Product & 51--200 \\
P8 & Process Automation & COO & 2--10 \\
P9 & Risk/Security Services & President & 2--10 \\
\bottomrule
\end{tabular}
\end{minipage}}
\hspace{3pt}
\adjustbox{valign=t}{\begin{minipage}{0.45\textwidth}
\centering
\textbf{(b) Public-Sector Participants}
\begin{tabular}{llll}
\toprule
ID & Agency Type & Role & Jurisdiction \\
\midrule
AP1 & City Mgmt & AI Program Manager & 500k+ \\
AP2 & City Mgmt & Finance Specialist & 20k+ \\
AP3 & Housing & Director of Innovation & 1M+ \\
AP4 & County Exec Office & Chief Privacy Officer & 1.5M+ \\
AP5 & City IT & Sr Manager / CDO & 300k+ \\
AP6 & City IT & Senior Manager & 300k+ \\
AP7 & City IT & AI/Data Fellow & 300k+ \\
AP8 & City IT & Privacy Specialist & 750k+ \\
AP9 & City IT & Director of IT & 95k+ \\
AP10 & Procurement & Procurement Specialist & 5M+ \\
\bottomrule
\end{tabular}
\end{minipage}
}
\end{table}

Our research design combines two complementary qualitative approaches. 
First, semi-structured interviews with vendors and public-sector practitioners explore how different stakeholders conceptualize the purpose, risks, and value of documentation tools such as the AI FactSheet. Second, we analyze a corpus of completed vendor FactSheets to understand how these documents function in practice and what types of information vendors disclose when responding to standardized prompts. This mixed qualitative design allows us to connect stakeholders’ stated goals and constraints with the realities of written disclosure artifacts.



\xhdr{Interviews} We conducted approximately 20 hours of semi-structured interviews with 19 participants involved in the creation, submission, evaluation, or use of AI FactSheets. 
Participants were recruited from organizations participating in the GovAI Coalition as well as government agencies evaluating or deploying AI systems (Table~\ref{tab:participants}). We aimed to capture perspectives across a range of professional roles and levels of technical involvement. Vendor participants included executives, product leaders, consultants, and engineers responsible for developing or marketing AI systems to government clients. Public-sector participants included procurement specialists, IT leaders, privacy officers, innovation staff, and program managers responsible for evaluating or overseeing AI-enabled systems.


Interviews lasted between 60 and 90 minutes and followed a semi-structured protocol designed to explore both participants’ experiences with the FactSheet and their broader views on AI documentation in procurement processes. 
Interviews were organized into four sections. First, participants described their professional roles and responsibilities related to AI systems or procurement decisions. Second, participants were asked to walk through their experiences creating, submitting, reviewing, or interpreting AI FactSheets. This portion of the interview often involved discussing specific sections of the document, and identifying challenges or uncertainties encountered during the disclosure process. Third, participants described the types of information they consider necessary to meaningfully evaluate AI systems during procurement. Finally, participants were asked to reflect on potential future approaches to documentation and information sharing, including certification schemes, disclosure incentives, and other governance mechanisms.

Interview transcripts were analyzed using inductive thematic analysis ~\citep{braun2006using} to identify recurring patterns in how stakeholders conceptualize documentation, disclosure risk, and evaluative needs. 
Following open coding, recurring concepts were grouped into higher-level themes that capture how participants understand the purposes and limitations of AI self-disclosure documentation. These themes informed the analytic focus of our subsequent document analysis.




\xhdr{FactSheet Analysis} 
Our second data source consists of 39 completed AI FactSheets submitted to the GovAI Coalition as of October 2025, including both publicly available and access-restricted documents. 
Each FactSheet contains responses to 23 standardized prompts (accessible online \citep{govai_factsheet}) that ask vendors to disclose basic descriptions about their system (``system overview''), development practices, evaluation methods, data sources, performance metrics, and governance processes.
Vendors respond to each prompt by writing in an open-response text box, granting vendors flexibility to decide how they will format or structure their responses.
Two research team members with expertise in AI governance and evaluation reviewed the corpus of completed factsheets.
We acknowledge that public-sector practitioners (\eg{} procurement professionals) may lack the necessary technical knowledge to make informed assessments of AI vendors' claims, as surfaced in past research ~\citep{johnson2025legacy}.
For these reasons, we evaluated the factsheets from the perspective of expert reviewers to assess whether they provided sufficient evidence to interpret their technical claims ~\citep{raji2022fallacy}: a necessary prerequisite for them to be useful to practitioners.
To analyze vendors' responses, we adopted a two-part analytic strategy.
First, we applied a structured rubric published by the Center for Democracy \& Technology  (``CDT'') ~\citep{cdt_transparency_rubric} to systematically score the amount of information in vendors' responses.
We complemented this sytematic analysis with open qualitative coding, described below.

\xhdr{Scoring factsheets using the CDT rubric} We applied selected dimensions of the CDT's \textit{Framework for Assessing Transparency in the Public Sector}~\cite{cdt_transparency_rubric} to assess patterns of disclosure related to system governance, evaluation practices, data use, and contextual limitations. 
We selected this framework because it provides one of the few publicly available rubrics designed specifically to evaluate transparency in AI systems used by public-sector organizations, and was published concurrently and independently of our team's research.

Because the CDT rubric was created to assess any form of vendor-provided AI documentation (\eg{} not necessarily in response to the GovAI factsheet), we first mapped each CDT rubric dimension to corresponding factsheet prompts that were most likely to contain relevant disclosures.
This mapping informed where (in which sections of the factsheet) our research team searched for relevant information.
For example, we mapped the CDT rubric's ``Evaluation and Testing'' dimension to factsheet prompts that asked vendors to disclose relevant ``Performance Metrics'' and results of any ``Independent Evaluations''.
Similarly, other rubric dimensions were applied only to those FactSheet prompts where the requested information aligned with the rubric’s criteria. 
Two members of the research team independently scored responses using the rubric’s three-point ordinal scale (minimal, partial, or substantive disclosure). Disagreements were resolved through discussion and consensus to produce a final agreed-upon score for each response.

\xhdr{Qualitative open coding} In addition to systematically scoring vendors' responses, we also conducted open coding of vendors' responses.
For each factsheet prompt (\eg{} ``\emph{Training Data}''), the two co-authors reviewed the full corpus of factsheet responses. From there, the authors grouped together related codes to characterize broader trends in the types of technical
information that vendors did (or did not) elect to disclose (\eg{} ``\emph{response does not disclose population training data was sampled from}''), or broader trends in how this information was presented (\eg{} ``\emph{displacement of responsibility due to reliance on third parties}'').
Once a codebook was created, the authors interpreted the trends we identified in how factsheets were completed, together with our findings about stakeholders' visions and goals for using the factsheet from semi-structured interviews.
This iterative analysis allowed us to place trends in how the factsheets were completed (\eg{} trends such as ``\emph{promotional language articulating alignment with public sector values}''), in conversation with our findings for how public sector workers were using them, and how vendors approached filling them out (\eg{} as an opportunity to communicate with prospective customers).

\xhdr{Integrating Methods \& Contextual Engagement} To answer our research questions, we iteratively revisited and refined our interpretations based on insights from interviews and the completed factsheets. This approach allowed us to identify three competing goals and visions for the factsheet that were shared \emph{across} both agency workers and vendors (Section \ref{sec:findings-stakeholders}), and broader, ecosystem-level obstacles to sharing information about public sector AI systems (Section \ref{sec:ecosystems}).

In addition to our formal data collection methods, our research team was embedded as participants and volunteers within the San José GovAI coalition throughout the project. The authors regularly attended coalition meetings, such as weekly ``industry relations'' meetings that convened vendors and public-sector purchasers. Although these meetings were not treated as study data, this embedded participation helped us develop a deeper understanding of the stakeholders our work sought to support, and the coalition’s broader goals. 

Because the research was conducted in partnership with coalition leadership, we also shared interim findings with members and invited practitioners to interpret the results. For example, we held a one-hour workshop at an in-person coalition summit where we presented preliminary study findings, and invited both vendors and public-sector participants to reflect on their implications. Insights from these engagements informed the final recommendations we developed for both individual practitioners, and the wider GovAI coalition.
\vspace{-5pt}
\section{Findings}
\begin{figure}
    \centering
    \includegraphics[width=0.8\linewidth]{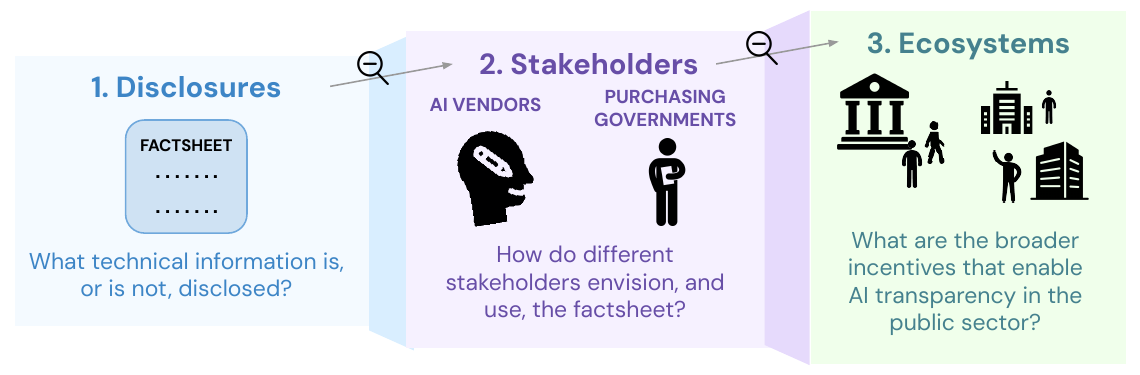}
    \caption{\textbf{Unpacking the GovAI FactSheet's role in facilitating meaningful disclosures}. Our findings begin by understanding what types of disclosures are currently being made, and then move outwards to examine the perspectives of key stakeholders (AI vendors and purchasing governments) and broader ecosystems of incentives that enable or disable meaningful disclosures about public sector AI systems. Understanding the broader incentives surrounding the FactSheet brings clarity to understanding how and why they were completed. }
    \label{fig:findings_diagram}
    \Description{A schematic showing technical disclosures feeding to stakeholder needs, feeding to ecosystem incentives and structure. This figure represents the nested interplay of FactSheets in public procurement.}
\end{figure}
In this section, we bring findings from our document analysis and interviews into conversation to examine how AI FactSheets are completed and used in public sector procurement. Rather than treating FactSheets as standalone artifacts, we analyze them as part of ongoing relationships between vendors and purchasing governments, as well as the broader public- and private-sector contexts in which these documents are produced and interpreted.



\subsection{Analyzing vendors' completed self-disclosures (RQ1)}

Across the full corpus of completed GovAI FactSheets, a consistent pattern emerges: while vendors generally provide relevant and coherent responses to each prompt, those responses frequently omit basic technical details about system design, training data, and performance. As a result, disclosures often gesture toward transparency without rendering systems legible to a reader seeking to understand how the AI was built or how it performs.

\begin{wrapfigure}{r}{0.5\linewidth}
    \includegraphics[width=\linewidth]{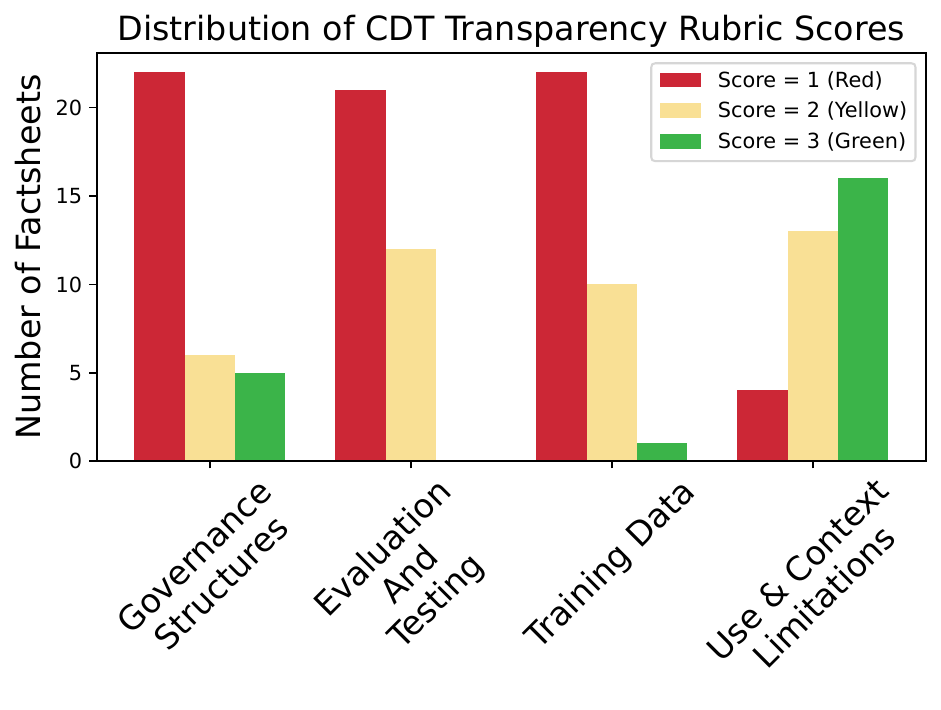}
    \caption{While vendors frequently provide meaningful information about intended use and contextual limitations, disclosures relevant to evaluation, governance, and training data remain sparse. This asymmetry underscores the mismatch between expectations that FactSheets function as evaluation tools and the realities of voluntary, public-facing documentation, reinforcing their practical role as relational, early-stage governance artifacts.}
    \label{fig:rubric_scores}
    \Description{A bar chart showing rubric scores for four categories: Governance structures, evaluation and testing, training data, and Use and Context limitations. For the first three categories, the majority of FactSheets received the lowest score, with only a few scoring higher. For the final category,more than half of the FactSheets received the highest score.}
\end{wrapfigure}

To characterize variation in disclosure quality across the dataset, we scored each FactSheet using the CDT’s rubric for assessing public-sector transparency. Figure~\ref{fig:rubric_scores} summarizes aggregate scores across four dimensions: use and context limitations, governance structures, training data, and evaluation and testing. Full rubrics and scoring criteria are provided in Appendix~\ref{app_methodology}, with example scoring tables shown in Table~\ref{tab:training_data_disclosures} and Table~\ref{tab:performance_metrics_disclosures}.

Vendors received the highest scores for use and context limitations
(16 of 39 scoring 3/3) 
and the lowest scores for training data and evaluation 
(22 of 39 scoring 1/3). 
The use and context dimension captures whether a FactSheet can ``\emph{clearly articulate what [the vendor’s] tool should be used for}.'' Many FactSheets offered clear descriptions of intended use, such as tools designed to support permitting workflows or detect urban blight. 
In contrast, disclosures about governance, training data, and performance were often sparse. 
In these dimensions, most vendors received the lowest possible score 
(22 of 39 scoring 1/3), 
indicating that ``little to no relevant information'' was disclosed. Even when explicitly prompted to share``\emph{any relevant information, links, or resources regarding your organization's responsible AI strategy}'', many vendors left fields blank, provided broken links, or stated that governance processes were still ``in progress.'' 
Because training data and evaluation are foundational to assessing bias, reliability, and fitness for use, 
we focus the remainder of this section on analyzing the variation and gaps in vendors’ disclosures along these two axes.

\xhdr{Unpacking vendors' training data disclosures}
The importance of training data curation in shaping system behavior is well established in the FAccT literature \cite{bender2018data,Gebru2021,pushkarna2022datacards}. In public-sector contexts, training data disclosures are a key site for assessing bias \cite{gerchick2023devil,hickok2024public}, validity \cite{coston2023validity}, and risks of dataset shift \cite{mulligan2019procurement}, particularly given vendors’ frequent reuse of off-the-shelf systems across jurisdictions \cite{veale2018fairness,kawakami2024studyingpublicsectorai}. These concerns are heightened by growing scrutiny of whether generative AI training data were obtained legally and with consent \cite{rueter2024want,johnson2025legacy}. 
We therefore examine what technical information vendors disclose about their training datasets in completed FactSheets.

Across the dataset, most responses lacked the information needed to understand what data were collected. 
This pattern is reflected in the large share of responses (22 of 39) receiving a rubric score of 1.
Vendors in this category offered only minimal or generic statements, such as that a dataset was ``collected with proprietary methods'' [FS8], that a model was ``trained by customer staff'' [FS9], or was trained using ``tens of thousands of images'' [FS10], without specifying the underlying data. 
Several disclosures relied on boilerplate language applicable to nearly any system, such as claims that models were trained on ``diverse, open-source datasets curated from the Internet'' [FS21].

A smaller subset of responses (10 out of 39) offered more insight into vendors' datasets, and thus were assigned a score of 2.
These disclosures typically identified the type of data used, such as a resident-facing chatbot trained on ``publicly available government webpages'' [FS36] (Table \ref{tab:training_data_disclosures}) or a permitting review tool trained on ``historical project files shared by the permitting departments of municipalities'' [FS5]. However, even these responses rarely specified how data were obtained or which populations they represented—for example, which municipalities contributed the records. As a result, the provenance and contextual relevance of many vendors' datasets remain difficult for practitioners to assess.

\xhdr{Unpacking vendors' reported  evaluations}
Performance evaluation is a central concern in public-sector AI deployment, where systems are often used in safety- or rights-critical contexts \cite{raji2022fallacy,pruss2023ghosting,adams2024no}. Prior work suggests that meaningful evaluations can help governments assess system capabilities and limitations while supporting more informed procurement decisions \cite{rubenstein2020federal,johnson2025legacy}. Accordingly, the GovAI FactSheet prompts vendors to report both how performance is measured and their current performance levels.

Vendors’ responses varied widely across domains and evaluation practices. To assess their informativeness, we adopt \citet{raji2021aiwideworldbenchmark}'s framework, which conceptualizes evaluation as comprising both a dataset and a metric, clarifying the information needed to interpret performance claims \cite{liao2021are}.

Across the dataset, over half of the FactSheets (21 out of 39) reported no performance results, receiving the lowest rubric score (Figure~\ref{fig:rubric_scores}).
While some vendors described performance concepts such as ``accuracy'' [FS28] or ``usefulness'' [FS1], these were rarely operationalized. A few referenced specific metrics (\eg{} BLEU or Word Error Rate) without reporting results (18 of these 21 referenced a metric abstractly, but contained no numbers). 
Others reported aggregate scores without describing the evaluation datasets, which were typically characterized only as ``internal''  [FS13] or ``proprietary'' data  [FS19] (\eg{} ``90\% response accuracy based on internal testing'' [FS7]).
None of the reported evaluations used publicly available datasets, limiting their interpretability and comparability.

\xhdr{Summary} This analysis provides one of the first empirical accounts of how AI vendors respond to structured self-disclosure requests in a public-sector procurement context. Although FactSheets are often framed as mechanisms for surfacing value-laden design choices \cite{govai_openletter}, many vendors disclosed little concrete information about training data or performance. Key details about data provenance, evaluation practices, and system behavior were frequently missing or underspecified, leaving important aspects of system capabilities and risks opaque. The consistency of these patterns across vendors motivates closer examination of the incentives and constraints shaping disclosure practices.
\vspace{-5pt}

\subsection{Understanding how stakeholders envision the purpose of AI FactSheet (RQ2)}\label{sec:findings-stakeholders}

Drawing on interview data, we analyze how participants describe their goals, expectations, and considerations when engaging with the FactSheet. We use examples from completed FactSheets to illustrate the orientations and priorities that interviewees described, rather than to evaluate whether those goals are achieved. Across interviews, participants articulated three recurring expectations they associated with the FactSheet: using it to showcase value, to support quality assessment, and to initiate longer-term relationships between vendors and government agencies.

\xhdr{FactSheets as showcasing value added}
Across interviews with vendors and public-sector practitioners, participants frequently described the FactSheet as an opportunity to showcase the value an AI system provides for a specific public-sector use case. Many participants described the FactSheet as a space for communicating benefits, differentiation, and practical relevance, rather than serving primarily as a vehicle for detailed technical disclosure.

When asked about their motivations for completing the FactSheet, vendors commonly framed it as a way to highlight what makes their system attractive in a competitive market.
One vendor emphasized wanting ``more opportunity to showcase what differentiates [our] product from the competition'' [VP3], while another described it as a way to gain ``any exposure, especially for small companies'' [VP1], explaining their desire to ``maximize the eyeballs'' that see it. These motivations were closely tied to the Coalition’s decision to make completed FactSheets public and searchable, positioning them as a marketplace where the FactSheet may serve as a potential customer’s first introduction to a product.
Some agency interviewees echoed similar priorities, emphasizing outcomes and problem-solving over technical detail.
As one public-sector participant explained, ``We don't want to get technical about how [the AI] works. I don't care about the programming, or what program you use, but how does it solve the problem?'' [AP2]

Consistent with these perspectives, vendors’ completed FactSheets frequently emphasized value and fit over substantive technical disclosure. Beyond describing intended use cases, responses often highlighted anticipated benefits such as efficiency, innovation, and alignment with public-sector values. Vendors used largely positive language, describing systems as able to ``\emph{revolutionize} government operations'' [FS4], ``help municipalities \emph{maximize efficiency}'' [FS29], or provide ``always-available, tireless, and patient support'' [FS16] (emphasis added). This orientation extended to assurances about system design and compliance, including claims that training relied on publicly available data [FS39], that updates occurred ``frequently'' [FS1], or broad commitments to ``thoughtful design [that] ensures you can use AI safely, responsibly, and ethically''  [FS3]. Together, these patterns suggest that many FactSheet responses may be structured to produce a favorable impression of vendors’ systems.

This perspective was also evident in how vendors interpreted underspecified prompts. In response to open-ended requests for performance evaluations, several vendors reported metrics emphasizing business value, such as a ``70 to 90 percent reduction in report-writing time'' [FS18] and reductions in task completion time to ``around 2–7 minutes'' [FS18].
While meaningful to prospective purchasers seeking to understand operational benefits, these metrics differ from the technical evaluations typically used to assess accuracy, bias, or robustness. Notably, these disclosures rarely included evaluations that would surface system limitations or potential harms.

\xhdr{FactSheets as quality assessment} A second major goal articulated by both governments and vendors was the use of FactSheets as tools for quality assessment. In this framing, documentation should support judgments about whether a system is fit for a particular public-sector context: whether it meets an agency’s technical, operational, legal, or responsible AI requirements, and whether its risks are acceptable given the intended use. 
This understanding aligns with the GovAI Coalition’s framing of the FactSheet as a procurement-stage tool. The Coalition's Open Letter describes the FactSheet as helping provide "evidence that the system fits our needs and is safe to implement in our jurisdictions" so that agencies can assess whether a system is appropriate for deployment, while not requiring the level of detail needed to reproduce or fully inspect it~\cite{govai_openletter}.
Participants consistently described quality assessment as a process that allows agencies to decide for themselves what is ``good enough,'' rather than relying solely on vendor assurances.

Agency participants emphasized the need for concrete technical and governance information, particularly around data provenance and responsible AI concerns.
One participant noted a persistent focus on training data, explaining: ``I’m always thinking about training data... and I know that this technology ethically puts women and people of color more at risk.''  [AP1] Others highlighted lifecycle risks, such as whether a system would continue to train on city data or how it might evolve over time.
Most of the vendors also expressed support for more rigorous evaluation, often framing it as a way to differentiate products and avoid poor-fit deployments.
At the same time, both groups acknowledged that agencies may lack the technical expertise or institutional capacity to interpret highly detailed disclosures, complicating the use of open-ended technical information.


As discussed above, however, completed FactSheets rarely provide the information needed to support meaningful quality assessment. While a small subset adopted an evaluation-oriented approach, offering specific metrics, describing testing regimes, or outlining governance workflows, most relied on high-level claims and often lacked baselines, contextual explanation, or decision-relevant implications.

For agencies attempting to use FactSheets to compare or assess systems, this vagueness limits the usefulness of FactSheets. Claims of ``high accuracy'' [FS29] are difficult to act on without knowing how accuracy was measured or on what data, while references to undefined ``edge cases'' [FS16] offer little guidance for anticipating real-world risk. Similar gaps appeared in disclosures about robustness, operating conditions, data dependencies, and governance practices. In these cases, FactSheets may signal that evaluation has occurred, but they do not provide enough substance for agencies to conduct their own assessments.

Agencies therefore often rely on external certifications such as SOC II or ISO standards as heuristics for baseline organizational practices. Yet, in the absence of comparable certification regimes for other dimensions of AI quality such as model behavior, bias, or contextual suitability, FactSheets depend heavily on voluntary technical disclosures. When those disclosures are sparse or ambiguous, the burden of quality assessment shifts back onto agencies that may lack the information, access, and resources to fill in the gaps.

\xhdr{FactSheets as building long-term relationships} 
Beyond showcasing value or supporting quality assessment, both vendors and agencies described a third, less explicit but widely shared goal: building trusted, long-term working relationships between governments and AI vendors. In this framing, the FactSheet is less valued for the completeness of its disclosures than for how it helps initiate and sustain a productive relationship over time in which better disclosures can happen. 
Participants described documentation as one component of an ongoing interaction that establishes baseline trust, signals good faith, and creates a shared starting point for future dialogue. 
From this perspective, the FactSheet is often treated as a starting point that invites iterative clarification, follow-up questions, and continued exchange, rather than as a final or self-sufficient account of the system.

This perspective emerged clearly in interviews emphasizing trust-building dynamics. 
For instance, several vendors linked openness to long-term opportunity rather than immediate transactional gain.
As one vendor reflected, ``the more you share, it’s been my experience in state and local government, the better opportunity you’ll have'' [VP1].  Vendors and agencies alike described FactSheets as living artifacts that evolve alongside procurement conversations or pilot deployments.
Agency participants echoed this sentiment, with one example describing how relational safety shaped disclosure practices: ``Because [the AI vendor] felt safe in the relationship with us, they knew it wasn't a `gotcha.' It was `do the best you can.' So they did, and their answer wasn't perfect'' [AP8]. 
Here, the FactSheet functioned not as a compliance artifact or evaluative gatekeeper, but as an entry point into collaborative problem-solving. The document enabled ongoing communication, allowing the agency and vendor to iteratively improve the system rather than treating early disclosures as final judgments.

Situating the FactSheet as one part of an ongoing relationship helps clarify how agencies can engage with disclosures as part of a continued exchange, rather than as final answers. In practice, government employees can treat underspecified responses to ask follow-up questions of vendors. For example, one vendor responded to a question about poor operating conditions by stating that outcomes ``totally depend on the use case''  [FS9], inviting agencies to participate in defining how they intend to apply the system before sharing more information about system performance. Similarly, another vendors' response that they ``manage [algorithmic bias] by a mandatory human review process''  [FS23], can provide a starting point for governments to ask more informed questions \eg{} about what these human reviews entail in practice. In this way, FactSheets can support knowledge sharing by establishing shared terminology, language, and reference points that can facilitate deeper disclosures as a procurement unfolds.

\vspace{-5pt}
\subsection{Participants' Views on Challenges and Structural Obstacles to FactSheet Efficacy (RQ3)}\label{sec:ecosystems}
In interviews, vendors and public-sector practitioners frequently described challenges that shaped what information could be shared through the FactSheet, independent of its specific prompts or design. Participants framed these challenges not as shortcomings of the FactSheet itself, but as reflections of broader structural conditions within the public–private AI procurement ecosystem in the United States.
This subsection presents participants’ own accounts of the legal, competitive, and organizational constraints that influence both vendors’ willingness and their ability to disclose information about the AI systems they sell to governments. 
In this section, we describe two recurring obstacles highlighted by vendors: first, concerns about information sensitivity in the context of public disclosure and competitive markets; and second, gaps in oversight introduced by increasingly complex AI supply chains. 

\xhdr{Information Sensitivity, Public Disclosure, and Competitive Risk}
Vendors consistently described legal and competitive constraints that shaped what they felt comfortable sharing. Intellectual property protection, competitive disadvantage, and the implications of FOIA or public records laws loomed large, particularly for smaller firms without extensive legal or compliance infrastructure.
As one vendor [VP1] explained, ``everything's FOIAed... what you're putting in could get right into your competitor's hands.'' 
The perception that written communication with government agencies may become public generated caution throughout the FactSheet completion process.

Participants emphasized that detailed technical disclosures, such as prompt engineering strategies, internal heuristics, or fine-tuning approaches, were viewed as core competitive assets.
One vendor [VP7] summarized this concern succinctly: ``The things I wouldn't want to put [on the FactSheet] would be our methods... that's our edge.'' Vendors also raised concerns that publicly available documentation could enable competitors to reverse-engineer system logic or infer architectural choices. In this context, self-disclosure is experienced not as a potential competitive liability. These concerns help explain why many FactSheet responses rely on generalized descriptions or business-oriented framing, even among vendors who expressed support for self-disclosure in principle.

\xhdr{Gaps in Visibility and Oversight Across the AI Supply Chain}
A related set of obstacles stems from the increasing complexity of AI supply chains. Many vendors do not build models end-to-end, instead integrating or fine-tuning foundation models developed by upstream providers. As a result, they often lack access to information about pretraining datasets, licensing arrangements, or model internals, even when agencies request such details. For systems built on APIs like GPT-4 or other foundation models, vendors frequently described an inability—not unwillingness—to disclose information they cannot access.
Vendors responded to these gaps in several ways. Some used the FactSheet to delineate the limits of their responsibility, emphasizing that performance depends heavily on downstream factors such as deployment context, data quality, or prompt design. 
Others positioned themselves explicitly as integrators rather than creators, highlighting a mismatch between what the FactSheet implicitly assumes (completers have full system visibility), and the realities of their role. Several participants suggested structural changes to the template, such as standardized selections for commonly used foundation models (as ``everyone is using GPT''). 

From a governance perspective, these dynamics raise fundamental questions about what self-disclosure can realistically look like when no single actor has full system-level knowledge. The FactSheet assumes centralized control and visibility that are increasingly rare. While vendors may be willing to be transparent, the structure of contemporary AI markets and the incentives facing upstream providers significantly constrains what can be disclosed, and by whom.
\vspace{-5pt}
\section{Discussion}
\vspace{-5pt}
Our findings challenge prevailing assumptions about what disclosure documentation can realistically achieve in public-sector AI governance. While FactSheets are often expected to function as standalone tools for evaluation, comparison, and oversight \cite{govai_openletter}, our analysis reveals structural limitations in relying on voluntary, publicly accessible vendor self-disclosures for these purposes. At the same time, we observe that FactSheets play an important role in practice by structuring early-stage dialogue between vendors and government agencies. Here, we examine the limits of documentation-based governance when disclosure is treated as a proxy for evaluative rigor, and propose an interpretive shift toward understanding such documentation as relational infrastructure. We then consider implications for AI governance design and future research in the FAccT community.
\vspace{-5pt}
\subsection{Unpacking the affordances and limitations of voluntary vendor self-disclosures}
Many disclosure initiatives assume that standardized documentation can function as a stand-alone artifact enabling meaningful assessment \cite{moss2021assembling,chang2022understanding,wright2024null,tang2025navigating}. In procurement contexts, these documents are often expected to support decisions such as comparing model offerings or informing purchasing choices \cite{govai_openletter,aaas_keyquestions,ieee_standard}. From this perspective, FactSheets are implicitly treated as evaluative tools.
But our findings suggest this expectation is difficult to realize in practice. Figure~\ref{fig:rubric_scores} highlights a pattern surfaced in both interviews and rubric-based assessment. While FactSheets often contain informative descriptions of intended use and contextual limitations, they provide far less information about evaluation practices, governance structures, and training data. This unevenness is not best understood as vendor non-compliance or a flaw in the GovAI template. Instead, it reflects a deeper mismatch between expectations placed on documentation and the structural conditions under which it is produced.

These constraints extend beyond the problems of open-ended prompts. Voluntary public disclosures are also shaped by legal risk, competitive pressure, upstream dependencies (\eg{} foundation models), and limited organizational visibility into system internals. Even when vendors complete FactSheets in good faith, these constraints limit what can be disclosed. As a result, documentation can create an appearance of disclosure and visibility without providing the verifiability that responsible governance often requires. The rubric scores reflect this dynamic, showing that dimensions most closely associated with evaluative rigor consistently receive lower scores across the dataset.

These limitations are not unique to the GovAI FactSheet. Rather, they reflect broader structural challenges facing governance regimes that rely on voluntary self-reporting as a proxy for evaluation. Similar dynamics may arise in other documentation-based governance approaches, including model cards, transparency reports, or regulatory self-assessment regimes such as those envisioned in emerging AI governance frameworks. Prior work has shown that voluntary industry disclosures tend toward selective and risk-averse reporting \cite{cheng2022how}. Our findings extend this insight by showing that in public-sector procurement, disclosures to government agencies are widely perceived by vendors as effectively public, intensifying concerns about competitive exposure and further constraining what is disclosed.


\xhdr{Implications}
We argue that based on the misaligned incentives we have discovered, the present AI FactSheet often is not completed in a way that can support the meaningful quality evaluation advocated for by FAccT scholars ~\citep{mulligan2019procurement,hickok2024public,johnson2025legacy}.
This finding has several immediate implications for public sector practitioners that presently rely on the factsheet.
First, we recommend to the GovAI Coalition that they do not base their quality certification solely on vendors' AI FactSheet responses. Instead, they should continue to use and emphasize the AI FactSheet as a helpful resource to scaffold early conversations with vendors.
GovAI could also explore and support complementary forms of assessment, such as enabling governments to share information with each other, \eg{} about their past experiences as purchasers in peer networks.

Second, we recommend to local governments (\eg{} procurement professionals) that they should not expect to use the AI FactSheet as the basis to support rigorous technical evaluation (\eg{} to assess adherence to their own responsible AI standards \citep{johnson2025legacy}). 
Because vendors are often incentivized to complete documentation in a manner that will showcase their product, we hypothesize that effective AI governance will likely require layered processes that combine documentation with complementary mechanisms such as independent audits \cite{costanza2022who,casper2024black}, sandbox pilots, or system testing.
We encourage the FAccT community to participate in creating meaningful pathways to support quality assessment that stand outside of, and can complement, vendor self-disclosures, such as infrastructure to support public sector workers in conducting their own independent evaluations ~\citep{kawakami2024situate,suresh2024participation,gosciak2026llms}.

We hypothesize that many of our findings about the limitations of voluntary self-disclosures may apply to many other governance regimes that ask AI companies to report technical details about their systems in response to open-ended prompts.
We urge policymakers who are considering structuring or basing a policy enforcement regime around (voluntary or mandatory) vendor self-disclosures to think critically about the potential shortcomings and disincentives that may exist in achieving meaningful transparency.
Future research can conduct more rich comparative analyses of how different governance regimes might combat some of these perverse incentives, by understanding whether different ways of eliciting documentation might influence (or discourage) more high-quality disclosures: for instance, such as understanding whether the ``mandatory'' nature of the EU AI Act incentivizes or discourages greater transparency ~\citep{EU_AI_Act_2024,golpayegani2023be}, or analyzing how vendors respond to the NYC LL 144's more explicit and specific requirements for how bias audits must be conducted (\ie{} using metrics such as the ``impact ratio'' ~\citep{wright2024null,gerchick2025auditing}).


\vspace{-5pt}
\subsection{Seeing self-disclosure documentation as a relationship, not compliance}
Disclosure documentation in public-sector AI governance is often implicitly treated as a compliance artifact: a document to be completed, submitted, and reviewed as evidence that disclosure obligations have been met \cite{cheng2022how,cdt_transparency_rubric,tang2025navigating}. 
Our findings suggest that this assumption does not align well with how FactSheets are produced or used in practice.
Instead, our interviews and observations suggest that the primary value of FactSheets lies in the relationships and processes they enable. In practice, FactSheets function less as static transparency documents and more as \emph{relational infrastructure} within which disclosure can happen. 
In this way, they act as coordination devices that structure dialogue between vendors, procurement officials, and other stakeholders during early procurement stages.

This emphasis on the continued relationships built between stakeholders builds upon \citet{bovens2007accountability}'s conceptualization of ``relational accountability'' as a continued relationship where an ``actor'' has an obligation to explain or justify their conduct to a ``forum''.
FAccT scholars have applied this conceptualization of relational accountability to analyze the relationships between AI vendors (developers, or ``actors''), purchasers, and members of the public who aim to hold them to account (the ``forum'') \citep{wieringa2020account,cobbe2021reviewable,cobbe23supplychains}.
This relational framing is also applicable to understand the role of documentation artifacts and infrastructure, which can be thought of as just one piece or component within a larger accountability relationship in which vendors are continuously accountable to purchasing governments throughout the lifecycle of public procurement ~\citep{mulligan2019procurement,rubenstein2020federal,johnson2025legacy}.

This framing helps explain why information about intended use, deployment context, and system boundaries is more consistently disclosed than deeper technical or evaluative details: these forms of information are often viewed as being less risky to share publicly and more immediately useful for determining fit and applicability.
Through this perspective, the primary value of the GovAI FactSheet lies not in its ability to resolve questions of quality, risk, or performance on its own, but in its capacity to initiate longer-term relationships in which those questions can be explored through additional channels. FactSheets can signal where deeper scrutiny is warranted, highlight areas of uncertainty, and provide a starting point for follow-up discussions, testing, or review. 

This interpretation resonates with prior work in the FAccT community that critiques transparency mechanisms when they become performative or decoupled from their intended recipients, and instead emphasizes the importance of acknowledging the larger relationships and power structures that transparency providers and recipients are embedded within \cite{birchall2021radical,disclosurebydesign22norval,corbett2023interrogating}. Importantly, reframing FactSheets as relational tools does not diminish the ambitions of initiatives like the GovAI Coalition. Rather, it clarifies how such efforts can succeed by not attempting to do everything at once. When FactSheets are positioned as conversation starters rather than evaluative endpoints, they better align with both vendor constraints and agency needs, and they avoid placing unrealistic expectations on a single, public-facing document.


\xhdr{Implications} Our call to view the AI FactSheet within the context of a continued relationship has several immediate implications for public sector practitioners.
We recommend that local governments approach the FactSheet understanding that it can serve as a starting point to guide continued conversations with vendors, even in scenarios where vendors' responses paint an incomplete picture of key technical design or evaluation decisions \citep{mulligan2019procurement,cdt_transparency_rubric}.
We encourage public sector practitioners to scaffold their surrounding procurement processes to explicitly encourage follow-up dialogue about vendors' FactSheet responses, clarification requests, and iterative exchanges of information towards the broader goal of establishing accountability within the vendor-purchaser relationship \citep{cobbe23supplychains}.
We believe that this paradigm shift towards understanding the role AI documentation plays \emph{within the context of existing stakeholder relationships} is a productive shift for the ACM FAccT community.
Specifically, we join past work \citep{disclosurebydesign22norval} in recommending that FAccT researchers pursue research about AI documentation in direct collaboration with the stakeholders who they hope will produce and consume documentation artifacts.
Our work contributes a detailed empirical account of how pre-existing systems of incentives in public sector procurement shape how documentation is produced, and also surfaces productive ways that completed FactSheets can be interpreted within these continued relationships.

\vspace{-5pt}
\section{Conclusion}
Our findings suggest that self-disclosure research and practice within the FAccT community would benefit from exploring how documentation artifacts function as embedded within governance systems, rather than evaluating them in isolation. 
Documentation artifacts like FactSheets, model cards, and impact assessments do not govern by themselves; they shape governance by structuring relationships, expectations, and institutional workflows.

By grounding this analysis in the lived practices surrounding the GovAI FactSheet, we hope to contribute not a critique of a particular initiative, but a reframing of how such efforts can be understood and strengthened. 
Recognizing and highlighting the relational role of FactSheets opens space for more realistic, constructive governance designs that combine documentation, incentives, and evaluative processes in ways that better reflect the realities of public-sector AI procurement.

\newpage
\section{Endmatter Statements}

\subsection{Generative AI Usage Statement}

During manuscript preparation, the authors used generative AI tools (specifically, ChatGPT, OpenAI, GPT-4–class models) for limited editorial support, including grammar and style editing, sentence-level clarity improvements, and assistance with structuring figures and tables. Generative AI tools were not used to generate original  content, analyses, arguments, findings, or interpretations. All academic contributions including study design, data collection, analysis, interpretation, and writing were produced by the authors. The authors take full responsibility for the originality, accuracy, and integrity of the manuscript.

\subsection{Ethical Considerations Statement}\label{sec:ethics}
To preserve anonymity of participating employees and cities, we assured interviewees that their participation was voluntary, they could decline to answer interviewer questions, and their responses would be kept anonymous. 
For sensitive or potentially identifying interview quotes, we exclude participant IDs to preserve anonymity. 
To mitigate the risk that participating cities are identified, we limit the amount of detail we provide about each cities' practices.
Participants' department and job titles were modified to reduce the risk of re-identification.

\begin{acks}

\end{acks}

\bibliographystyle{ACM-Reference-Format}
\bibliography{refs}

\newpage
\appendix

\section{Methodology}
\label{app_methodology}
This appendix provides a detailed account of the study design, data collection, and analysis procedures summarized in Section~\ref{methodology} of the main paper. We report full interview protocols, recruitment and access considerations, coding procedures, rubric application details, and methodological limitations that are omitted from the main text for space.


To examine how AI self-disclosure documentation is produced and used in practice, we employed three complementary methods: semi-structured interviews, qualitative analysis of completed AI FactSheets, and contextual engagement through a multi-stakeholder workshop. This appendix provides procedural detail and analytic justification for each method.

\subsection{Interview Methodology}
We began by conducting 20 hours of semi-structured interviews with key stakeholders involved in the submission and use of AI FactSheets within the GovAI Coalition. Our goal was to explore how the different actors interact with and conceptualize the purpose and value of self-disclosure documentation, and how these expectations shape their approach to the FactSheet itself. 

\subsubsection{Recruitment}
We recruited participants from both vendor organizations participating in the GovAI coalition and government agencies that use or evaluate AI systems. Participants represented a range of roles, such as CEOs, product managers, IT managers, data scientists, procurement officials, and financial officers. 
Our recruitment process aimed to capture a diverse set of employees’ perspectives on the AI FactSheet, across multiple cities and vendors. As shown in Table~\ref{tab:participants}, most agency participants worked in technology-focused roles in their city’s IT or Innovation departments. Additionally, most vendors were involved in the development or decision making for their organizations. Participants in both groups included leaders who made decisions on behalf of their department or company, and workers whose day-to-day responsibilities involved managing AI technologies.
We began the study by reaching out to potential participants who were frequently and publicly involved in the GovAI Coalition meetings and working groups. From this group, we intentionally focused on inviting participants that represented a wide range of regions and maturity surrounding AI (e.g., whether or not they had adopted any public-facing AI policies). Although this approach has limitations, we learned that establishing trust through shared connections was important for government employees and AI vendors to feel comfortable speaking openly with academic researchers. 

\subsubsection{Data Collection}
Following Veale et al., we adopted a semi-structured interview approach to allow for flexibility in discussions. This allowed participants to spend more time discussing the phases of the procurement process that were closest to their role and expertise. The interviews ranged from 60 to 90 minutes and our protocol included four sections.  As a baseline, we asked participants about their current role and work responsibilities relating to AI. We defined “AI” for participants using the OECD definition of “any machine-based system that can make predictions, recommendations, or decisions”, and provided examples of qualifying systems. We chose the OECD definition to be intentionally inclusive of a wide set of predictive technologies beyond the contemporary focus on generative AI and LLM-based chatbots.  We then focused on participants’ experiences in creating, submitting, reviewing, or using AI FactSheets. We asked participants to walk through how they approach using the FactSheet, identifying pain points and overall impressions as they went. In this section, we also discussed the motivations that stakeholders might have in approaching the FactSheet as an information sharing tool. 
Then, we asked participants to highlight their needs in using or assessing FactSheets. In this section, we discuss the participant’s expertise, needs, and concerns in using a self-disclosure documentation tool for procurement. 
Lastly, we asked participants about the types of incentives they would like to see included in future approaches to self-disclosure documentation and information sharing. This included new forms of documentation, markets for information sharing, badging systems for certification, and other potential solutions. 
The study was approved by a university IRB. We include our complete interview protocol in Appendix XX, and discuss ethics and participant safety considerations in Section XX. 

\subsubsection{Analysis}
We conducted inductive thematic analysis of the interview transcripts using an interpretive, comparative approach. Following open coding, we identified recurring concepts and patterns in how participants described the functions of the FactSheet and the barriers to its effective use. These codes captured both practical and normative dimensions of participants’ experiences, such as efforts to signal credibility, concerns about disclosure risk, and divergent expectations of what information sharing should achieve. 

Building on these patterns, we grouped the themes into higher-level categories that correspond to the main functions we identified across interviews. This allowed us to map how participants’ goals and constraints shape the use cases for self-disclosure documentation. 

\subsection{FactSheet Analysis Methodology}
In addition to the semi-structured interviews, we analyzed the contents of completed vendor FactSheets submitted to the GovAI coalition. This collection of FactSheets included both publicly available and access restricted FactSheets. Our goal was to understand how these documents function as tools of self-disclosure and communication across the intended uses we identified in our interviews. Specifically, we asked: How do FactSheets perform, or fail to perform, the goals that participants hold: showcasing system values, enabling quality assessment, and supporting early and consistent dialogue?

\subsubsection{Further Framing}
We approached the FactSheets as artifacts of communication rather than as objective disclosures of system properties. Because we cannot infer vendors’ intentions from the text alone, we adopted the perspective of a hypothetical public sector reader and examined how each FactSheet might function in each of the possible goals. For each document and each field response, we answered questions such as:
Does this FactSheet effectively communicate the benefits or potential value of the system?
Does it provide sufficient information to enable meaningful quality or risk assessment?
Does it provide enough shared language to initiate a productive conversation between a vendor and an agency?

This approach allowed us to assess the functional use of the FactSheet within the framework of the broader analysis of audience and author mismatches. 

\subsubsection{Coding Procedure}
We first conducted exploratory open coding on the full set of vendor FactSheet responses. Each FactSheet contains responses to 23 standardized questions covering areas such as system overview, model development, training and test data, performance, and governance practices. We used this inductive approach to identify recurring trends in how vendors responded to each prompt. 

These codes captured features such as rhetorical or persuasive techniques (e.g., “positive sales language,” “emphasis on client success stories,” “references to trust or fairness without evidence”), as well as descriptive indicators of technical disclosure (e.g., “no information about training data provided,” “partial metric reporting,” “specific model architecture named”).

After open coding, we mapped each type of response to the different goals that the FactSheet attempts to serve. For instance, “positive sales language” often supports the showcasing function, while "no information about training data provided" might be a barrier to meaningful quality assessment in many contexts; but it may still enable the FactSheet to be used as a way to build long term relationships.

\subsubsection{Applying the CDT Transparency Assessment Rubric}\label{apdx:methods-cdt}
To complement our qualitative analysis, we conducted a structured assessment of vendor disclosures using \textit{A Framework for Assessing Transparency in the Public Sector}, a rubric recently proposed by researchers at the Center for Democracy \& Technology (CDT). The rubric was designed to evaluate the self-disclosure of AI systems procured or used by public-sector entities and is explicitly motivated by documentation-based disclosure practices similar to the GovAI Coalition’s FactSheet. To our knowledge, this study represents the first empirical application of the CDT rubric to a corpus of real, completed vendor self-disclosure documents.

\textit{Mapping rubric dimensions to FactSheet content} The CDT rubric defines multiple dimensions of transparency, including governance structures, evaluation and testing, training data and methodologies, and use and context limitations. Because the GovAI FactSheet is a multi-question document rather than a one-to-one instantiation of the rubric, we did not score entire FactSheets holistically for each rubric dimension. Instead, we mapped each rubric category to the specific FactSheet questions most likely to elicit relevant information.

For example, to assess the Evaluation and Testing dimension, we examined vendor responses to FactSheet questions corresponding to (1) internal performance evaluation and (2) independent or third-party evaluation. Similarly, other rubric dimensions were applied only to those FactSheet prompts where the requested information plausibly aligned with the rubric’s criteria. This approach reflects our analytic goal of assessing whether vendors provided meaningful evaluative information when given the opportunity to do so, rather than penalizing them for omissions in unrelated sections.

Within each mapped set of responses, we considered the substance of all relevant statements collectively. That is, scores reflect whether, taken together, vendors’ responses meaningfully addressed the rubric criteria, rather than whether any single question response was sufficient in isolation.

\textit{Scoring procedure and reliability} Two members of the research team independently scored each relevant FactSheet response using the CDT rubric’s three-point ordinal scale (1 = minimal or insufficient disclosure; 2 = partial or moderate disclosure; 3 = substantive disclosure). Scoring focused on the specificity, clarity, and evaluative usefulness of the information provided, in line with the rubric’s original intent.

Overall, the research team observed high agreement in initial scoring. Where disagreements occurred, they primarily stemmed from differing interpretations of rubric thresholds (e.g., what constituted “partial” versus “substantive” disclosure). To resolve these cases, the coders met to discuss discrepancies and jointly refine operational definitions for each score level. These refinements were applied consistently across the dataset to arrive at a final, agreed-upon score for each rubric dimension and FactSheet.


\textit{Role of the rubric within the broader analysis} We emphasize that the CDT rubric assessment was not intended to serve as a definitive audit or certification of vendor practices. Rather, it provides a structured, comparative lens on patterns of disclosure across FactSheets and serves as a complement to our qualitative, open-coded analysis. In the findings, we use aggregated rubric scores to contextualize and corroborate themes identified through interviews and close reading of FactSheet responses, particularly with respect to uneven disclosure across different self-disclosure dimensions.



\subsubsection{Interpretation}
Our final analysis combines both the qualitative insights and descriptive summaries to evaluate how effectively the FactSheets fulfill different goals of self-disclosure. We used these FactSheet insights to help validate the interview findings by confirming some of the most relevant phenomena discussed. We use illustrative excerpts from vendor responses to demonstrate how particular rhetorical strategies or omissions shape the document’s utility for different readers.

We would also like to note that this paper represents a subset of our broader coding framework. The full codebook includes additional categories and subthemes, but we report only those findings most relevant to understanding audience-author mismatches and the multi-purpose nature of disclosure documentation in AI procurement.



\subsection{Methodological Limitations}
As with any qualitative, interpretive study of governance practices, our methodological approach involves trade-offs. Below, we outline key limitations associated with our interview data, qualitative FactSheet analysis, and application of the CDT transparency rubric, and describe how these limitations shape the scope of our claims.

\subsubsection{Limitations of the interview sample}
Our interview participants were drawn from vendors and public-sector agencies engaged with the GovAI Coalition. While this allowed us to study a rare and difficult-to-access population—particularly private vendors selling AI systems to government—it also means that our findings reflect the perspectives of actors who have already opted into a self-disclosure-oriented governance initiative. Vendors or agencies that choose not to participate in coalitions or voluntary disclosure efforts may face different incentives or constraints that are not fully captured here.

Additionally, recruitment relied in part on existing relationships and visible participation within coalition activities. This trust-based recruitment strategy was necessary to facilitate candid discussion of sensitive topics such as disclosure risk, competitive pressure, and legal concerns, but it may have skewed participation toward individuals who are more engaged in governance conversations. As a result, our findings should be interpreted as illuminating how self-disclosure is negotiated among relatively motivated actors, rather than as a representative survey of all public-sector AI vendors or agencies.

Finally, interview data reflects participants’ self-reported experiences and perceptions. While these accounts provide essential insight into motivations, fears, and expectations that shape documentation practices, they may not fully reflect organizational behavior beyond the individual level.

\subsubsection{Limitations of qualitative FactSheet coding}
Our qualitative analysis of completed FactSheets involved interpretive open coding of narrative responses. This approach necessarily involves researcher judgment, particularly when identifying rhetorical strategies (e.g., promotional framing) or assessing the level of technical specificity in disclosures. While we mitigated this limitation through iterative coding, comparison across documents, and triangulation with interview data, the analysis does not claim objectivity or exhaustiveness.

Importantly, we do not infer vendors’ intentions from the text of FactSheets. Our analysis is explicitly reader-centered, asking how a hypothetical public-sector reader might interpret the information provided and what purposes those responses could plausibly serve. As such, our findings characterize the functional affordances of documentation rather than the motivations or good faith of individual vendors.

In addition, our analysis focuses on a subset of themes most relevant to understanding the multi-purpose nature of FactSheets and challenges to meaningful self-disclosure. The full range of information contained in the documents is broader than what we report here.

\subsubsection{Limitations of the CDT rubric application}
The CDT transparency rubric provides a structured framework for assessing disclosure quality, but it was not originally designed as a scoring instrument for multi-question narrative documents like the GovAI FactSheet. As a result, applying the rubric required interpretive mapping between rubric dimensions and relevant FactSheet questions. While this mapping was done systematically and transparently, alternative mappings or interpretations are possible.

Rubric scoring also involves normative judgments about what constitutes “sufficient” or “substantive” disclosure, particularly in the absence of standardized benchmarks or shared expectations across vendors. Although we used multiple coders and consensus-based reconciliation to improve reliability, the resulting scores should be understood as indicative patterns rather than definitive evaluations of individual systems or vendors.

Finally, because the rubric emphasizes evaluative and governance-oriented disclosures, lower scores may reflect structural constraints on public, voluntary documentation rather than poor self-disclosure practices per se. This limitation aligns with—and indeed motivates—our broader argument that documentation-based governance tools face inherent constraints that should be accounted for in policy design.

\subsubsection{Scope of claims}
Taken together, these limitations mean that our findings are best understood as interpretive and diagnostic rather than prescriptive or evaluative. We do not claim that FactSheets “fail,” nor that vendors act in bad faith. Instead, our analysis identifies recurring structural tensions and incentive misalignments that shape how self-disclosure documentation functions in practice. These insights are intended to inform future governance design, research, and experimentation, rather than to serve as compliance judgments or performance rankings.
\section{Interviews}
\label{app_interviews}
\subsection{Vendor Interviews}
We began the interview by reminding the participant of our informed consent protocol (approved by our institution’s IRB board), and asking for their consent to record.

\subsubsection{Part 1: Introduction and Background}
\textit{The goal is to understand the company’s product, the participant's specific role, and their personal relationship to AI procurement. Understand any prior government procurement interactions.}

The goal of this interview is to learn more about your experiences with local government procurement specifically for artificial intelligence, or AI, technologies. We would like to start with a review of the GovAI factsheets, trying to understand the challenges and issues you face, as well as hearing about your needs and perspectives. Then we will broaden the scope to the incentives and challenges in working with local governments overall. For this conversation, we adopt a wide definition of AI as "any machine-based system that can make predictions, recommendations, or decisions". This would include technologies such as large language models, predictive analytics systems, resume screening technology, ChatGPT, etc.

\textbf{Q1.0:} Can you describe your company and the core AI products or services?
Probes: Company information (location, size), types of technology

\textbf{Q1.1:} Can you tell me a bit about your current role in the company? How do you interact with government agencies in your current role?
Have you worked with government procurement in the past?
What are the challenges you have experienced in those interactions? (An unbiased insight from the vendors before being prompted with Factsheets, should include some probes just in case)
Probes: communications, expectations, work load, lack of ROI, etc
On the other hand, what are some of the benefits or incentives you have experienced in those interactions? (An unbiased insight from the vendors before being prompted with Factsheets, should include some probes just in case)
Probes: communications, expectations, visibility, work load, ROI, etc

\textbf{Q1.2:} Do you have an idea of what percent of your company's (or your organization's) business comes from local and state governments?

\textbf{Q1.3:} Has your company ever filled out the GovAI factsheet? Another organization’s factsheet? 
Have you ever gone through the process of sharing your product’s performance for a government procurement of an artificial intelligence technology?
\begin{itemize}
    \item If YES: How were you involved? How did it go? Which products/projects?  Why did you fill out the factsheet?
    \item If NO: Has your company ever considered or talked about procuring AI?
What has prevented engagements in the past?
\end{itemize}

\subsubsection{Part 2: Factsheet Probe}
\textit{The goal is to gather some first reactions to the Factsheet as a procurement artifact. }

For the next part of our interview, we would like to understand your organization’s perspective and approach to Factsheets and other procurement artifacts. 

\textbf{Q2.0:} Familiarization/First Impressions. Are you familiar with the Factsheet?
What are your overall impressions or reactions to Factsheets?
What motivated you to fill one out?
Do you have any clarifying questions before we walk through it together?

\textbf{Q2.1:} Which part or parts of this document, or similar documents, do you have the most experience in addressing?

\textbf{Q2.2:} Walkthrough. Can you briefly walk us through how your organization has filled out or would typically go about filling out an artifact like this? 
Who might be involved? And how long would it take to appropriately fill out the Factsheet?
We would like to try and find out the work that is involved in answering [an area or question of interest] section from a vendor’s perspective. (Emphasize trying to understand, not calling out)
Think about sections that should be easy in theory, but end up more difficult than expected in practice. 
We noticed that some vendors struggled to answer questions about [a specific section or question]. I was wondering if you had any insight as to why those types of questions might be difficult to answer.
(We pick some questions and/or we let them pick some sections)

Are there any sections that might be challenging to answer or that you have difficulty in understanding?
Which ones?
Are there any sections that might be burdensome or irrelevant for your product?
Which ones?
Are there any sections that are especially useful in presenting information about your product?
Which ones?
Are there any sections that you think should be adjusted?	
Which ones, and how?

\textbf{Q2.3:} Other documentation tools. Have you ever used other documentation tools for procurement in different situations?
Can you describe those documentation processes?

\textbf{Q2.4:} Other evaluation tools. Have you ever used other evaluation tools for procurement in different situations?
Can you describe those evaluation tools or processes?

\subsubsection{Part 3: Factsheet as Facilitator}
\textit{The goal is to understand how the Factsheet influences vendors’ decisions to interact and ultimately bid on contracts.}

For the next part of our interview, we would like to explore how documentation and reporting requirements might influence your organization’s decision to bid on a contract. 

\textbf{Q3.1:} Do you anticipate documentation requirements like the Factsheet might deter your organization from pursuing a local government contract? If so, why?
Probe if needed- things like a lack of return on investment, not enough people or time, unclear requirements?

\textbf{Q3.2:} On the other hand, do you believe documentation requirements like the Factsheet can benefit you in your relationship with local government or help your organization in any way?
Probes if needed: do they build trust, speed up evaluation, or in other ways improve decision making through the procurement process? 
Are there any specific sections that you believe are especially helpful for you or the procurement team?
If so, how would these sections help review teams better understand your product?

\subsubsection{Part 4: Incentive Discovery}
\textit{The goal is to understand the participant's needs and desires to identify the incentives that would make Factsheet completion worthwhile, or the incentives that would drive vendors toward [interaction, contract bidding, and sharing of information].}

For the last part of our interview, we'd like to understand your opinions and wishes for improving incentives for AI procurement.

What kind of [information] do you typically provide to prospective customers about your AI system, and in what way? Why do you provide this information in this way?
Who in your organization is responsible for giving this information, and in what way?
What motivates you to provide this information?
What makes it challenging to provide information (be transparent) about AI systems?
What do you believe would help make providing this information more worthwhile?

\textbf{Q4.1:} What do you believe would help make completing the Factsheet more worthwhile (or help increase your likelihood of competing for contracts, or interacting with local government procurement processes)?
Probe - certifications or badges, preferred listing in a marketplace, faster contract processing, etc
Do you have any experiences where other procurement processes involved similar incentives?
Can you describe those processes and the incentives?

\textbf{Q4.2:} How do you believe a badge based incentive tier would help make completing the Factsheet more worthwhile (or help increase your likelihood of competing for contracts, or interacting with local government procurement processes)?
Probe - By a badge based incentive tier, we are picturing a system where a score is given based on the quality of the factsheet. That score translates to a certain badge level, and that badge has certain incentives associated with it. Higher badge levels would result in better incentives, and lower badge levels, lower incentives. 

\textbf{Q4.3:} Are there any additional support measures that would help lower barriers to participation in Coalition AI initiatives and local government procurement?
Probe - advisory/consultation, technical assistance, financial compensation(?), etc

\textbf{Q4.3:} Can you imagine any new resources that could help you address these challenges?
What resource format would be most helpful?
ex: Checklists? Templates? Trainings?

The GovAI Industry relations committee aims to build a marketplace where amazing vendors like you! who participate in responsible AI initiatives can be recognized (rewarded) to potential customers (local governments). 
Can you think of ways we could design the marketplace that would be beneficial to your company?

\subsubsection{Part 5: High Level Solutions}

We would also like to try and capture some high level ideas for what potential solutions might look like. 

\textbf{Q5.1:} If you could redesign the Factsheet or procurement process, what would be your top changes?

\textbf{Q5.2:} Are there any other artifacts or processes that you think would help improve your engagement?

\subsection{Agency Interviews}
We began the interview by reminding the participant of our informed consent protocol (approved by our institution’s IRB board), and asking for their consent to record.

\subsubsection{Part 1: Background}
\textit{The goal is to understand the agency’s mission, the participant's specific role, and their personal relationship to AI procurement. Understand any prior procurement interactions.}

The goal of this interview is to learn more about your experiences with local government procurement specifically for artificial intelligence, or AI, technologies. We would like to start with a review of the GovAI factsheets, trying to understand the challenges and issues you face, as well as hearing about your needs and perspectives. Then we will broaden the scope to the incentives and challenges in working with vendors in procurement overall. For this conversation, we adopt a wide definition of AI as "any machine-based system that can make predictions, recommendations, or decisions". This would include technologies such as large language models, predictive analytics systems, resume screening technology, ChatGPT, etc.

\textbf{Q1.0:} Can you describe your agency’s/team’s main mission and responsibilities?

\textbf{Q1.1:} Can you tell me a bit about your current role, and any past work experiences or responsibilities relating to artificial intelligence procurement?

\textbf{Q1.2:} Have you ever been involved in a past procurement of an artificial intelligence technology?
\begin{itemize}
    \item If YES: How were you involved? Which products/projects? 
    \item If NO: Has your [agency] ever considered or talked about procuring AI? What has prevented engagements in the past?
\end{itemize}

\subsubsection{Part 2: Factsheet Probe}
\textit{The goal is to understand how the Factsheet is used and understood internally. }

For the next part of our interview, we would like to understand your organization’s perspective and approach to Factsheets and other procurement artifacts. 

\textbf{Q2.0:} Familiarization/First Impressions. Are you familiar with the Factsheet?
What are your overall impressions or reactions to Factsheets?
Do you have any clarifying questions before we walk through it together?

\textbf{Q2.1:} Which part or parts of this document, or similar documents, do you have the most experience in addressing?

Think about - does your org have a mature AI procurement process? What does that look like? Etc. 

\textbf{Q2.2:} Walkthrough. Can you briefly walk us through how your team would typically evaluate an artifact like this? 
Who might be involved? And how long would it take to appropriately assess the vendor based on the Factsheet?
We noticed that some vendors struggled to answer questions about [a specific section or question]. I was wondering if you had any insight as to why those types of questions might be difficult to answer.
What kinds of answers would you find useful?

Are there any sections that might be challenging to assess or that you have difficulty in understanding?
Which ones?
Are there any sections that might be burdensome or irrelevant for your agency?
Which ones?
Are there any sections that are especially useful in presenting information about a vendor’s product?
Which ones?
Are there any sections that you think should be adjusted?	
Which ones, and how?

\textbf{Q2.3:} Other tools. Have you ever used other documentation tools for procurement in different situations?
Can you describe those documentation processes?

\subsubsection{Part 3: Capacity and Needs}
\textit{The goal is to understand how the Factsheet influences agencies’ decisions to interact with vendors.}

For the next part of our interview, we would like to discuss how the Factsheet might influences your decisions to interact with vendors.

\textbf{Q3.1:} (If not discussed earlier) At what stage of the workflow do Factsheets appear in your assessment process?
Do you believe they create bottlenecks?
Do you believe they speed up screening or evaluation?

\textbf{Q3.2:} Do you have the in-house expertise you need to understand and evaluate each of the sections of the Factsheet?
If so, can you tell us about what roles and expertise those individuals have?
If not, what support would be helpful?
Probe - Consultations, training, etc

\textbf{Q3.3:} Have you ever experienced/observed a vendor being deterred by the Factsheet requirement?
Please explain

\textbf{Q3.4:} On the other hand, do you believe documentation requirements like the Factsheet can benefit vendors in their relationship with local government or help them in any way?

\subsubsection{Part 4: Incentive Discovery}
\textit{The goal is to discover ways that the GovAI group can attract additional vendors and tailor requirements. }

For the last part of our interview, we'd like to understand your opinions and wishes for improving incentives for AI procurement.

\textbf{Q4.0:} Can you think of any ways to adjust the process or Factsheet so that AI products with different levels of risk can be assessed differently?
Example: tiered risk levels for pothole detection vs facial detection vs recidivism prediction

\textbf{Q4.1:} What do you believe would help make completing the Factsheet more worthwhile for vendors (or help increase their likelihood of competing for contracts, or interacting with local government procurement processes)?
Probe - certifications or badges, preferred listing in a marketplace, faster contract processing, etc
(if relevant) Do you have any experiences or knowledge of other procurement processes that involved similar or other incentives?

\textbf{Q4.2:} One concept that has been discussed previously is a “GovAI Stamp of Approval” or badging system. What documentation requirements or thresholds would you suggest that vendors meet to obtain that stamp?
Likely need to explain the concept of the stamp in more detail, and give potential probing suggestions
Probe - By a badge based incentive tier, we are picturing a system where a score is given based on the quality of the factsheet. That score translates to a certain badge level, and that badge has certain incentives associated with it. Higher badge levels would result in better incentives, and lower badge levels, lower incentives. 

\textbf{Q4.3:} Are there any additional support measures that you could provide to help lower barriers to participation in local government procurement?
Probe - advisory/consultation, technical assistance, financial compensation(?), etc

\textbf{Q4.4:} Can you imagine any new resources that could help you address these challenges?
What resource format would be most helpful?
ex: Checklists? Templates? Trainings?

\subsubsection{Part 5: High Level Solutions}
We would also like to try and capture some high level ideas for what potential solutions might look like. 

\textbf{Q5.1:} If you could redesign the Factsheet or procurement process, what would be your top changes?

\textbf{Q5.2:} Are there any other artifacts or processes that you think would help improve vendor engagement?

\newpage
\section{Training Data Disclosures}
\begin{table}[H]
\centering
\small
\begin{tabular}{p{0.08\linewidth} p{0.32\linewidth} p{0.52\linewidth}}
\toprule
\textbf{Score} & \textbf{Disclosure Characterization} & \textbf{Example Response} \\
\midrule
\multicolumn{3}{p{\linewidth}}{
\textbf{Training Data Disclosures.} 
\textit{How was the AI system trained? What data was used? How often is data added to the training set?}
} \\
\midrule
1 &
The vendor does not provide information about how its system was designed and trained and does not provide training data information. &
\textit{Example:} "(Urban object detection) Our company personnel curated our proprietary training datasets using both in-house and commercial datasets. These datasets were collected with proprietary methods, and have been collected and trained legally and exceeding industry best-practice standards." \\
\midrule
2 &
The vendor provides high-level information that may allow procurers or other stakeholders to draw inferences, but does not provide specifics. &
\textit{Example:} "(Resident-facing chatbot) We train our AI on information that our government customers have already published on their websites. Training data includes webpages and information about relevant communication channels." \\
\midrule
3 &
The vendor provides specific information about the training methodologies and data. &
\textit{Example:} "(AI writing support) Our service uses individual free (non-business) user accounts for product improvement and training purposes. Users' text data is included in a random sample of content that is de-identified, sanitized, and anonymized." \\
\bottomrule
\end{tabular}
\caption{Illustrative examples of real vendors' responses to training data disclosure prompts, categorized using the three-point rubric (reproduced here) from the CDT \cite{cdt_transparency_rubric}.}
\label{tab:training_data_disclosures}
\end{table}

\section{Performance Metrics Disclosures}

\begin{table}[H]
\centering
\small
\begin{tabular}{p{0.08\linewidth} p{0.42\linewidth} p{0.42\linewidth}}
\toprule
\textbf{Score} & \textbf{Disclosure Characterization} & \textbf{Example Response} \\
\midrule
\multicolumn{3}{p{\linewidth}}{
\textbf{Performance Metrics Disclosures.} 
\textit{What are the performance metrics? What is your current level of performance on these metrics? How can the user monitor performance when deployed?}
} \\
\midrule
1 &
The vendor does not provide information about its testing approach or provides reassurances without specifics. &
\textit{Example:} (Internal productivity chatbot) ``Overall, [the system] demonstrates high accuracy based on internal validation tests and from current customer feedback.'' \\
\midrule
2 &
The vendor provides some information about how it tested its systems, such as results of a performance evaluation but no methodology. &
\textit{Example:} (Internal productivity chatbot) ``We measure precision: whether the solution provides answers that are not only correct and accurate but also aligned with what the question specifically asks for.
 Current: 79.5\% Target: 90\%' \\
\midrule
3 &
The vendor provides detailed information about how it tested its system, including methodology and results, who conducted those evaluations, and when and how it updates and re-tests. &
No responses in the dataset received a score of $3$. \\
\bottomrule
\end{tabular}
\caption{Illustrative examples of real vendors' responses to performance metric disclosure prompts, categorized using the three-point rubric (reproduced here) from the CDT \cite{cdt_transparency_rubric}.}
\label{tab:performance_metrics_disclosures}
\end{table}

\end{document}